\documentclass[usenatbib]{mn2e}

\usepackage{graphicx}
\usepackage{graphics}
\usepackage{amsmath}
\usepackage{amssymb}
\usepackage{color}
\usepackage{url}
\usepackage{hyperref}
\usepackage{pdflscape}
\usepackage{ulem}
\usepackage{xcolor}

\newcommand{\begit}{\begin{itemize}}
\newcommand{\enit}{\end{itemize}}
\newcommand{\begen}{\begin{enumerate}}
\newcommand{\enen}{\end{enumerate}}

\newcommand{\beq}{\begin{equation}}
\newcommand{\eeq}{\end{equationf}}
\newcommand{\beqa}{\begin{eqnarray}} 
\newcommand{\eeqa}{\end{eqnarray}}

\def\aap{A\&A}

\def\apj{ApJ}
\def\apjl{ApJL}
\def\apjs{ApJS}

\def\mnras{MNRAS}
\def\nat{Nature}

\def\araa{ARAA}

\begin{document}

\title[Characteristic Momentum of Hot Galactic Winds]{The Characteristic Momentum of Radiatively Cooling Energy-Driven Galactic Winds}
\author[Lochhaas, Thompson, \& Schneider]{Cassandra Lochhaas$^{1,2}$, Todd A.~Thompson$^{2}$, Evan E.~Schneider$^{3,4}$\\
$^{1}$ Space Telescope Science Institute, 3700 San Martin Drive, Baltimore, MD, 21218, USA\\
$^{2}$ Department of Astronomy and the Center for Cosmology and Astro-Particle Physics, Ohio State University, \\
140 West 18th Avenue, Columbus, OH 43210, USA\\
$^{3}$ Department of Physics and Astronomy, University of Pittsburgh, 3941 O'Hara St, Pittsburgh, PA, 15260, USA\\
$^{4}$ Department of Astrophysical Sciences, Princeton University, 4 Ivy Lane, Princeton, NJ 08544, USA}

\maketitle

\begin{abstract}
Energy injection by supernovae may drive hot supersonic galactic winds in rapidly star-forming galaxies, driving metal-enriched gas into the circumgalactic medium and potentially accelerating cool gas. If sufficiently mass-loaded, such flows become radiative within the wind-driving region, reducing the overall mass outflow rate from the host galaxy. We show that this sets a maximum on the total outflow momentum for hot energy-driven winds. For a spherical wind of Solar metallicity driven by continuous star formation, $\dot p_\mathrm{max} \simeq 1.9\times10^4\ M_\odot\ \mathrm{yr}^{-1}\  \mathrm{km\ s}^{-1}(\alpha/0.9)^{0.86}\left[R_\star/(300\ \mathrm{pc})\right]^{0.14}[\dot M_\star/(20\ M_\odot\ \mathrm{yr}^{-1})]^{0.86},$ where $\alpha$ is the fraction of supernova energy that thermalizes the wind, and $\dot M_\star$ and $R_\star$ are the star formation rate and radius of the wind-driving region. This maximum momentum for hot winds can also apply to cool, ionized outflows that are typically observed in starburst galaxies, if the hot wind undergoes bulk radiative cooling or if the hot wind transfers mass and momentum to cool clouds within the flow. We show that requiring the hot wind to undergo single-phase cooling on large scales sets a minimum on the total outflow momentum rate. These maximum and minimum outflow momenta have similar values, setting a characteristic momentum rate of hot galactic winds that can become radiative on large scales. We find that most observations of photoionized outflow wind momentum fall below the theoretical maximum and thus may be signatures of cooling hot flows. On the other hand, many systems fall below the minimum momentum required for bulk cooling, indicating that perhaps the cool material observed has instead been entrained in or mixed with the hot flow.
\end{abstract}


\section{Introduction}

Galactic outflows, driven by feedback from massive stars, directly impact galaxy formation. Without star formation feedback, galaxies produce too many stars \citep{Keres2009,Murray2010,Bower2012,Ceverino2014,Hopkins2014} and are too metal-rich \citep{Tremonti2004,Erb2006,Finlator2008,Peeples2011}. While galactic winds are observed in many low- and high-redshift galaxies, the driving mechanism is not fully understood --- supernovae, stellar winds, radiation pressure, magnetic fields, and cosmic rays may all contribute to driving galactic outflows, or one mechanism may dominate for galaxies with specific properties \citep[for reviews, see][]{Veilleux2005,Heckman2017}. 

Winds are observed to be multiphase; hot winds are tracked by X-ray emission \citep{Cappi1999,Strickland2000,Strickland2004,Strickland2007,Li2011,Yukita2012,Lopez2020}, cooler ionized and atomic gas is observed primarily in UV/optical absorption lines \citep{Heckman2000,Rupke2005,Grimes2009,Steidel2010,Martin2012,Rubin2014,Heckman2015,Chisholm2017}, broad nebular emission lines \citep{Erb2012,Newman2012,Arribas2014,Davies2019,ForsterSchreiber2018}, or via resonant scattering \citep{Martin2013}, and the coldest components are observed in molecular transitions \citep{Walter2002,Sturm2011,Leroy2015,GonzalezAlfonso2017,Fluetsch2019}. Different physical driving mechanisms may dominate for different wind phases, and theoretical considerations can constrain different potential physical explanations/mechanisms by direct comparison to data \citep{Coker2013,Zhang2014,Krumholz2017,Buckman2020}.

X-ray emitting hot winds naturally arise when over-pressurized, thermalized gas expands. Supernovae provide an energy source for heating the gas and likely contribute significantly to the driving of hot winds from some rapidly star-forming galaxies \citep{CC85,Heckman1990}. Cooler high-velocity components traced by UV/optical absorption against the stellar continuum or molecular emission are much more difficult to explain. Cool gas in outflows may arise due to entrainment of cool gas within a hot wind (e.g., \citealt{Cooper2009,Scannapieco2015,Banda-Barragan2016,Bruggen2016,Schneider2017,Zhang2017}), may be a signature of the hot winds undergoing bulk radiative cooling to low temperatures in a single phase on large scales outside the driving region \citep{Wang1995a,Wang1995b,Silich2003,Thompson2016b,Lochhaas2018,Schneider2018}, or may reflect momentum and energy transfer via mixing between the hot and cool phases \citep{Gronke2018,Fielding2020,Gronke2020,Kanjilal2020,Li2020,Schneider2020,Tan2020}. Other physical mechanisms like radiation pressure on dust and cosmic rays have also been suggested. 

\citet[][hereafter CC85]{CC85} show how a constant injection of energy and mass into a spherical region produces hot, pressurized gas that expands outward to drive a supersonic, hot wind. \citet{Silich2003,Silich2004,TenorioTagle2007,Wunsch2007,Wunsch2008,Wunsch2011} build on this model by including the radiative cooling of the wind inside the injection region, which may inhibit the wind that escapes the driving region. The ability of the wind to cool depends strongly on its mass loading, the ratio of the mass outflow rate to the star formation. The critical mass loading necessary for cooling to set in within the wind-driving region was derived in \citet{Lochhaas2017} in the context of the second generation stars seen in globular clusters \citep[see][for the critical wind luminosity for cooling]{Wunsch2007}. For values of the mass loading larger than the critical mass loading, larger fractions of the wind-driving volume radiatively cool and less of the injected energy and mass escapes the system as a wind. This critical mass loading implies upper limits on the mass outflow rate and wind momentum rate of a hot wind, which can limit the ability of winds to affect their environment. 

Observations of outflows can constrain wind momentum rates through estimates of the mass outflow rate and observed outflow velocity \citep{Rupke2005,Heckman2015,Heckman2016,Chisholm2017}, allowing for a direct comparison between the limits expected from theory and observed systems. However, such observations focus on cool-warm gas detectable in optical and UV absorption. In theories where the hot wind dominates the dynamics of the cool outflow --- e.g., because the cool gas is precipitated directly from the hot gas, is directly ram-pressure accelerated, or is accelerated by mixing with the hot flow --- we can derive bounds on the observed cool gas momentum rate from the physics of the hot flow, or vice versa.

In this paper, we first show that there is a maximum of the hot wind momentum injection rate due to radiative cooling in the interior wind-driving region for highly mass-loaded flows. This reduces the mass and energy that escapes the wind-driving region, effectively self-limiting the asymptotic wind kinetic power and force. If the hot gas dominates the dynamics of the cool gas, which is the case if the cool gas is accelerated by or precipitated directly from the hot gas, the maximum on the momentum and energy of the hot wind translates to a maximum on the cool outflow momentum and energy as well. In order for the hot wind to cool monolithically in a single phase \citep[e.g.][]{Thompson2016b}, the wind must be sufficiently mass-loaded, and this minimum mass loading factor required for single-phase cooling on large scales is of the same order of magnitude as the maximum mass loading rate for a hot wind we derive. Together, the maximum and minimum momentum injection rates for hot winds provide a benchmark for interpreting observations in the context of any model of cool gas outflows where the dynamics are ultimately controlled by the hot phase.

In \S\ref{sec:betacrit}, we derive the maximum injected mass loading factor and the associated maximum wind momentum rate. In \S\ref{sec:min}, we derive the minimum wind momentum of hot winds under the requirement of single-phase cooling on large scales. In \S\ref{sec:char}, we combine the maximum and minimum to produce a characteristic wind momentum rate and explore how it varies with the parameters of the problem. We compare these theoretical maxima and minima to observed values of UV-absorbing cool winds in \S\ref{sec:obs} and discuss other outflow models that can enhance wind momentum rates above the theoretical maximum or produce cool outflows below the theoretical minimum in \S\ref{sec:discussion}. We conclude in \S\ref{sec:summary}. 

\section{Critical Mass-Loading and Maximum Momentum}
\label{sec:betacrit}

We consider a picture in which supernovae inject mass and energy into a spherical region of radius $R_\star$. Following CC85, we assume the supernovae thermalize to produce a hot region of gas that then undergoes adiabatic expansion, becoming a galactic wind. We write the mass deposition rate of the wind within the injection region as
\begin{equation}
\dot M_\mathrm{wind} = \beta\ \mathrm{SFR} \label{eq:Mdot}
\end{equation}
where $\beta$ is the mass loading factor of the wind, and SFR is the star formation rate. We parameterize the energy deposition rate of the wind within the injection region as
\begin{equation}
\dot E_\mathrm{wind} = \alpha \dot E_\mathrm{SN} = 3\times10^{41}\ \mathrm{ergs\ s}^{-1}\ \alpha\ \mathrm{SFR} \label{eq:Edot}
\end{equation}
where $\alpha$ is the thermalization efficiency of the supernovae, i.e. the fraction of supernova energy $\dot E_\mathrm{SN}$ that enters into the energy of the wind-driving region. Equation~(\ref{eq:Edot}) assumes that $\alpha\ 10^{51}$ ergs is deposited by each supernova, and that a supernova occurs every 100 years for a SFR of $1\ M_\odot$ yr$^{-1}$.

Based on previous work, we expect that when $\beta$ is large enough, the interior of the injection region will cool radiatively. To estimate this critical value of the mass-loading parameter, $\beta_\mathrm{crit}$, above which we expect a larger and larger portion of the driving region to be radiative, we set the cooling time equal to the advection time $t_\mathrm{cool}=t_\mathrm{adv}$ at $r\rightarrow0$. Taking 
\begin{align}
t_\mathrm{cool}&=\frac{3}{2}\frac{P}{\Lambda(T)n^2} \label{eq:tcool_max} \\
t_\mathrm{adv}&=r/v \label{eq:tadv}
\end{align}
and using the expressions from CC85 for the density, pressure, and velocity in the limit $r\ll R_\star$, near the inner core of the injection region,
\begin{align}
P(r\ll R_\star)&=\frac{0.118 \beta^{1/2}\ \mathrm{SFR}^{1/2} \alpha^{1/2} \dot E_\mathrm{SN}^{1/2}}{R_\star^2} \label{eq:P} \\
n(r\ll R_\star)&=\frac{0.296 \beta^{3/2}\ \mathrm{SFR}^{3/2}}{\mu m_p R_\star^2 \alpha^{1/2} \dot E_\mathrm{SN}^{1/2}} \label{eq:n} \\
v(r\ll R_\star)&=\frac{0.269 \alpha^{1/2}\dot E_\mathrm{SN}^{1/2} r}{R_\star \beta^{1/2}\ \mathrm{SFR}^{1/2}}, \label{eq:v}
\end{align}
we find that 
\begin{align}
t_\mathrm{cool}&=\frac{2.02 \alpha^{3/2} \dot E_\mathrm{SN}^{3/2} R_\star^2 \mu^2 m_p^2}{\Lambda(T(r\ll R_\star))\beta^{5/2}\ \mathrm{SFR}^{5/2}} \label{eq:tcool_2} \\
t_\mathrm{adv}&=\frac{3.717\beta^{1/2}\ \mathrm{SFR}^{1/2} R_\star}{\alpha^{1/2} \dot E_\mathrm{SN}^{1/2}}. \label{eq:tadv_2}
\end{align}
In the above expressions, $P$ is the gas pressure, $n$ is the number density, $\Lambda(T)$ is the radiative cooling function\footnote{In this paper, we consider the cooling function for fully ionized gas only. Supernovae may seed dust production within the wind injection region, enhancing the radiative cooling efficiency \citep[e.g.,][]{Draine1981,Dwek1987,Martinez-Gonzalez2016}. However, dust is also rapidly sputtered and destroyed in hot gas, reducing the efficiency of dust as a coolant \citep{Scannapieco2017}. Here, we consider purely thermally-driven winds, in which a hot, uniform-temperature plasma fills the wind injection region and is responsible for driving the wind, so it is unlikely for dust to drastically alter the cooling function before it is destroyed.} for a temperature $T$, $R_\star$ is the radius of the wind-driving region, $\mu$ is the mean weight per particle, which we take to be $\mu=0.6$ for a fully ionized gas, $m_p$ is the proton mass, and the temperature within the injection region is given by the ideal gas law as
\begin{equation}
T(r \ll R_\star)=\frac{0.399\mu m_p \alpha \dot E_\mathrm{SN}}{k_B \beta\ \mathrm{SFR}} \label{eq:T}
\end{equation}
where $k_B$ is the Boltzmann constant. Assuming a cooling function of the form
\begin{equation}
\Lambda(T)\approx \Lambda_0 \left(\frac{T_0}{T}\right)^{0.7} 
\,\,(10^5 < T < 10^{7.3}\,{\rm K})
\label{eq:Lambda}
\end{equation}
with $\Lambda_0=1.1\times10^{-22}$\,ergs cm$^3$ s$^{-1}$ and $T_0=10^6$\,K at Solar metallicity, we find that the critical mass loading parameter is given by\footnote{\citet{TenorioTagle2007} derived a similar critical threshold for cooling within a wind-driving region by integrating the full set of hydrodynamics equations, rather than using the CC85 assumptions as done here. They expressed their threshold in terms of a critical wind luminosity, rather than a critical mass-loading factor. If we convert our $\beta_\mathrm{crit}$ into a wind luminosity, we obtain similar values to those found by \citet{TenorioTagle2007}.}
\begin{equation}
\beta_\mathrm{crit}^{3.7}=0.284 \frac{\mu^{2.7}m_p^{2.7}}{k_B^{0.7}T_0^{0.7}}\frac{\alpha^{2.7}\dot E_\mathrm{SN}^{2.7} R_\star}{\Lambda_0\ \mathrm{SFR}^{3.7}}. \label{eq:beta_crit}
\end{equation}
For fiducial parameters of $\alpha=0.9$, $\mu=0.6$, $R_\star=300$ pc, and $\mathrm{SFR}=20 M_\odot$ yr$^{-1}$ to represent a compact starburst or a dense star forming region in a high-redshift galaxy, we find that
\begin{align}
&\beta_\mathrm{crit}\simeq0.77 \nonumber \\
&\times\left(\frac{\alpha}{0.9}\right)^{0.73}\left(\frac{\mu}{0.6}\right)^{0.73}\left(\frac{R_\star}{300\,\mathrm{pc}}\right)^{0.27}\left(\frac{\mathrm{SFR}}{20 M_\odot\,\mathrm{yr}^{-1}}\right)^{-0.27}. \label{eq:beta_crit_max}
\end{align}
For this value of $\beta_\mathrm{crit}$, the central temperature inside the wind injection region is $T\sim10^{7.2}$ K (equation~\ref{eq:T}), which justifies our use of equation~(\ref{eq:Lambda}) as an approximation to the cooling function. Larger values of $\beta$ produce cooler central temperatures.

\begin{figure}
\centering
\includegraphics[width=\linewidth]{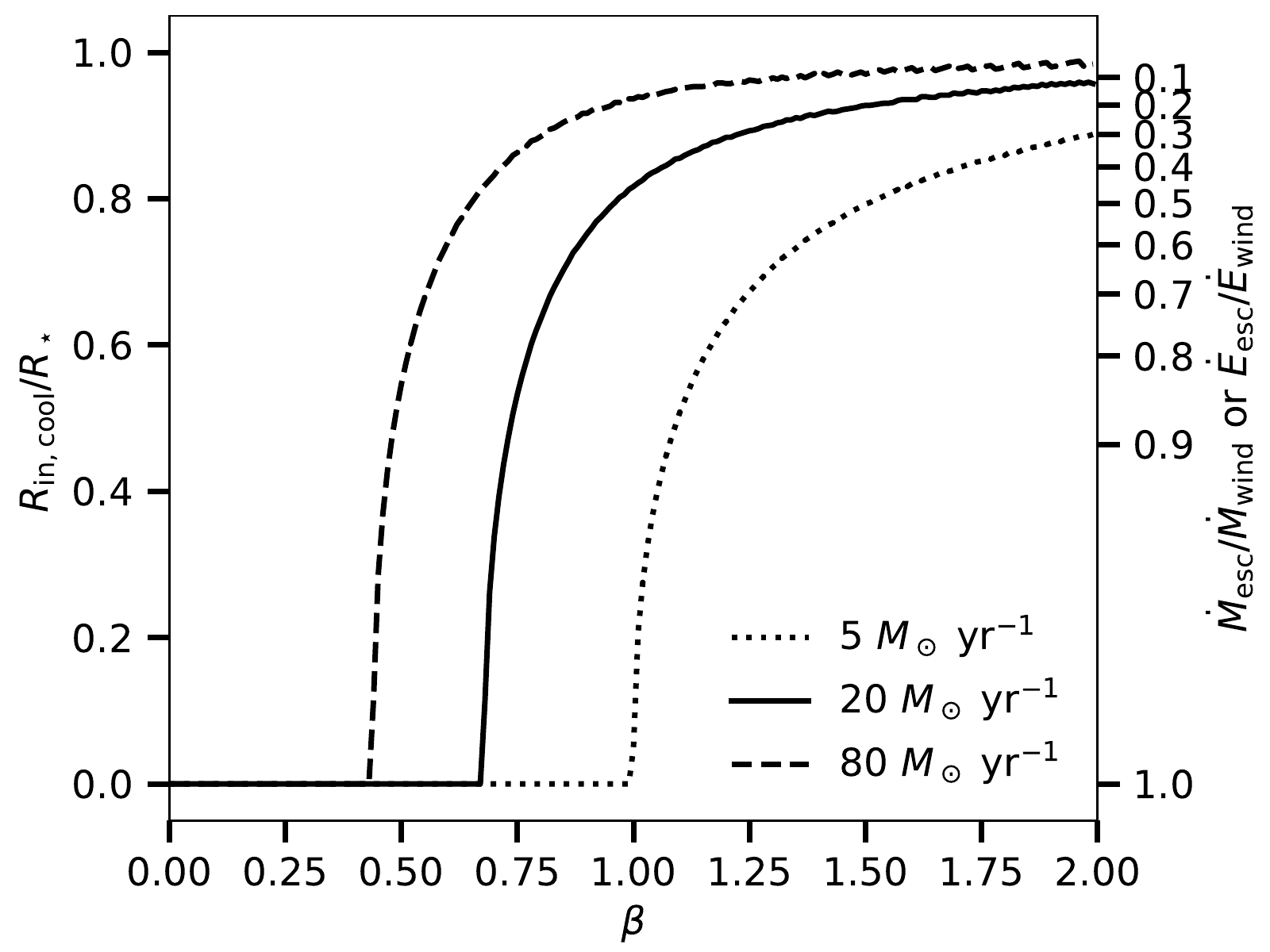}
\caption{The inner cooling radius $R_\mathrm{in,cool}$ in units of $R_\star$ (left axis) and the ratio of escaping mass to injected wind mass, $\dot M_\mathrm{esc}/\dot M_\mathrm{wind}$ or the equivalent ratio of escaping energy to injected wind energy, $\dot E_\mathrm{esc}/\dot E_\mathrm{wind}$ (right axis), as functions of mass loading factor $\beta$ for three values of SFR. As $\beta$ increases, the fraction of the wind-driving region volume that cools increases at first drastically, but never quite reaches unity so the escaping wind mass and energy never quite reach zero. The minimum $\beta$ at which $R_\mathrm{in,cool}/R_\star>0$ is $\beta_\mathrm{crit}$ (equation~\ref{eq:beta_crit_max}). As SFR increases, $\beta_\mathrm{crit}$ decreases because the density in the wind-driving region is larger, promoting cooling at smaller $\beta$. These curves were calculated with $\alpha=0.9$, $\mu=0.6$, $R_\star=300$ pc, and the three values of the SFR of $5$ (dotted), $20$ (fiducial, solid), and $80$ (dashed) $M_\odot$ yr$^{-1}$.}
\label{fig:Rcool}
\end{figure}

We define $R_\mathrm{in,cool}$ as the radius within which the wind material cools and is retained within the injection region, which is dependent on $\beta$, SFR, and $\dot E_\mathrm{wind}$. In order to find $R_\mathrm{in,cool}$ for a given value of $\beta$, we use an iterative process. We compute the radius within the cluster where $t_\mathrm{cool}=t_\mathrm{adv}$, using the radial profiles from CC85 and the full form of the cooling function from \citet{Wiersma2009}. Then, we adjust the normalization of the density profile of the material outside of $R_\mathrm{in,cool}$ to be lower by a factor of $1-(R_\mathrm{in,cool}/R_\star)^3$, following our assumption that none of the material within $R_\mathrm{in,cool}$ escapes, and recalculate $R_\mathrm{in,cool}$ using the new density profile with the lower normalization. We iterate on this process until $R_\mathrm{in,cool}$ converges. Figure~\ref{fig:Rcool} shows the value of $R_\mathrm{in,cool}/R_\star$ computed in this way for different values of $\beta$, for our fiducial parameters. When $\beta<\beta_\mathrm{crit}$, there is no cooling in the wind-driving region so $R_\mathrm{in,cool}=0$. $\beta_\mathrm{crit}$ can be read off of Figure~\ref{fig:Rcool} as the minimum $\beta$ where $R_\mathrm{in,cool}>0$, which is $\beta_\mathrm{crit}\approx0.67$ for fiducial parameters. The small difference between $\beta_\mathrm{crit}$ calculated in this way and in equation~(\ref{eq:beta_crit_max}) is due to the small difference between the full cooling function and the analytic approximation to the cooling function assumed in equation~(\ref{eq:Lambda}).

\begin{figure}
\centering
\includegraphics[width=\linewidth]{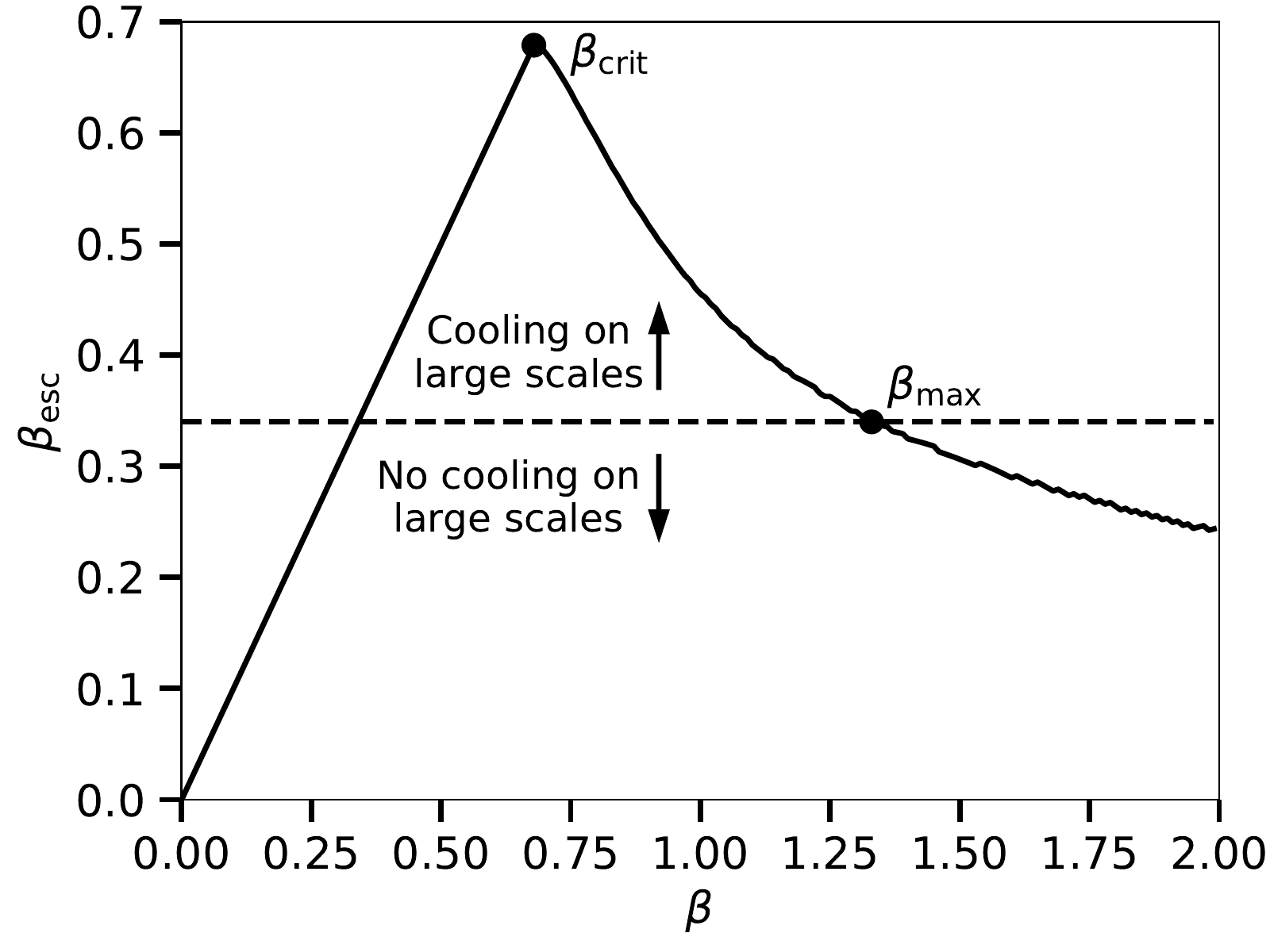}
\caption{The relationship between $\beta_\mathrm{esc}$, the mass loading factor of only that wind that escapes the wind driving region, and $\beta$, the mass loading factor of the injected wind within the wind driving region. The maximum value of $\beta_\mathrm{esc}$ occurs at $\beta_\mathrm{crit}$ and is marked with a point (see \S\ref{sec:betacrit}). Also shown as the dashed horizontal line is the minimum value of $\beta_\mathrm{esc}$ that allows single-phase radiative cooling of the wind on large scales. Where these curves cross is the maximum value of $\beta$ that will allow single-phase cooling (see \S\ref{sec:min}) and is marked with a point labeled $\beta_\mathrm{max}$. Note that for $\beta\leq\beta_\mathrm{crit}$, $\beta_\mathrm{esc}=\beta$ exactly. This figure shows the relation only for fiducial parameters of SFR =$20\ M_\odot$ yr$^{-1}$, $\alpha=0.9$, $R_\star=300$ pc, and $\mu=0.6$.}
\label{fig:betaesc}
\end{figure}

We assume that only wind deposited in $R_\mathrm{in,cool} < r < R_\star$ can escape the wind-driving region and contribute to the escaping large-scale wind. The escaping wind mass, $\dot M_\mathrm{esc}$, is given by
\begin{equation}
\dot M_\mathrm{esc}=\beta_\mathrm{esc}\ \mathrm{SFR} \label{eq:Mdot_esc}
\end{equation}
where $\beta_\mathrm{esc}$ is the ``effective" mass loading factor of only the wind that can escape the wind-driving region, and is given by
\begin{equation}
\beta_\mathrm{esc}=\beta\left[1-\left(\frac{R_\mathrm{in,cool}}{R_\star}\right)^3\right]. \label{eq:beta_esc}
\end{equation}
We define an analogous $\alpha_\mathrm{esc}$ and $\dot E_\mathrm{esc}$ as well. $\dot M_\mathrm{esc}$ and $\dot E_\mathrm{esc}$ are the mass and energy outflow rates of only the wind that actually leaves the wind driving region, while $\dot M_\mathrm{wind}$ (eq.~\ref{eq:Mdot}) and $\dot E_\mathrm{wind}$ (eq.~\ref{eq:Edot}) are the mass and energy deposition rates of all supernovae exploding within the cluster and all material they sweep up within the injection region. Thus $\dot M_\mathrm{esc}/\dot M_\mathrm{wind}=\dot E_\mathrm{esc}/\dot E_\mathrm{wind}=\beta_\mathrm{esc}/\beta=\alpha_\mathrm{esc}/\alpha$ is simply the volume fraction of the cluster that does not radiatively cool, which is $1-(R_\mathrm{in,cool}/R_\star)^3$; this relation holds because we assume homogenous mass and energy deposition within the injection region. Because no part of the injection region cools unless $\beta>\beta_\mathrm{crit}$, $\dot E_\mathrm{esc}/\dot E_\mathrm{wind}$ and $\dot M_\mathrm{esc}/\dot M_\mathrm{wind}$ are constant and equal to one for $\beta<\beta_\mathrm{crit}$. The right-side axis in Figure~\ref{fig:Rcool} shows the ratios $\dot M_\mathrm{esc}/\dot M_\mathrm{wind}=\dot E_\mathrm{esc}/\dot E_\mathrm{wind}$. Figure~\ref{fig:betaesc} shows the relationship between $\beta_\mathrm{esc}$ and $\beta$ for fiducial parameters. If, for example, the SFR is $20M_\odot$ yr$^{-1}$ and $\beta=1$, implying that just as much mass enters the wind as mass that forms stars, then $\dot M_\mathrm{wind}=20 M_\odot$ yr$^{-1}$. In this case, the inner region of the starburst with $r<R_\mathrm{in,cool}$ radiatively cools, retaining a fraction of this wind within the cluster and reducing the amount of wind driven out of the cluster. From Figures~\ref{fig:Rcool} and~\ref{fig:betaesc}, $\dot M_\mathrm{esc}/\dot M_\mathrm{wind}\sim0.5$ and $\beta_\mathrm{esc}\sim0.5$, implying that only $10M_\odot$ yr$^{-1}$ of wind material successfully escapes the cluster.

Because the amount of escaping wind increases with increasing $\beta$, but then decreases for $\beta>\beta_\mathrm{crit}$, the maximum mass outflow rate and thus the maximum momentum rate of the wind occurs for $\beta=\beta_\mathrm{crit}$. We define the momentum rate of the escaping wind as
\begin{equation}
\dot p_\mathrm{esc}=\dot M_\mathrm{esc} v_\mathrm{wind}, \label{eq:pdot}
\end{equation}
where $\dot M_\mathrm{esc}$ is given by equation~(\ref{eq:Mdot_esc}) and $v_\mathrm{wind}$ is given by
\begin{equation}
v_\mathrm{wind}=\left(2\frac{\dot E_\mathrm{esc}}{\dot M_\mathrm{esc}}\right)^{1/2}. \label{eq:vwind}
\end{equation}
$v_\mathrm{wind}$ is the asymptotic wind velocity for $r\gg R_\star$. The wind velocity is larger for larger $\alpha$ and is smaller for larger $\beta$, but has no dependence on the fraction of the wind that escapes the wind-driving region because $\dot E_\mathrm{esc}$ and $\dot M_\mathrm{esc}$ both contain the same factor of $1-(R_\mathrm{in,cool}/R_\star)^3$, so $\dot E_\mathrm{esc}$ and $\dot M_\mathrm{esc}$ can be replaced by $\dot E_\mathrm{wind}$ and $\dot M_\mathrm{wind}$ in equation~(\ref{eq:vwind}). Note that the unchanging value of $v_\mathrm{wind}$ regardless of how much of the injection region cools is predicated on our assumption that the mass and energy injection, as well as any additional material providing the mass loading, are all distributed uniformly throughout the injection region. This idealized assumption is a limitation of our model.

The maximum of the escaping wind momentum is then given by
\begin{equation}
\dot p_\mathrm{max,esc} = \beta_\mathrm{crit}\ \mathrm{SFR}\ v_\mathrm{wind}. \label{eq:pdotmax_betacrit}
\end{equation}
Scaling this to our fiducial parameters, we find that

\begin{align}
&\dot p_\mathrm{max,esc} = 1.9\times10^4\ M_\odot\ \mathrm{yr}^{-1}\ \mathrm{km\ s}^{-1} \nonumber \\
&\times\left(\frac{\alpha}{0.9}\right)^{0.86}\left(\frac{\mu}{0.6}\right)^{0.36}\left(\frac{R_\star}{300\ \mathrm{pc}}\right)^{0.14}\left(\frac{\mathrm{SFR}}{20\ M_\odot\ \mathrm{yr}^{-1}}\right)^{0.86}. \label{eq:pdot_max}
\end{align}

The expression for $\beta_\mathrm{crit}$ (equation~\ref{eq:beta_crit}) can be generalized to any constant source of mass and energy, $\beta\dot M$ and $\alpha\dot E$, within an injection region $R_\star$ assuming the gas metallicity is solar:
\begin{equation}
\beta_\mathrm{crit}^{3.7} = 0.284\frac{\mu^{2.7}m_p^{2.7}}{k_B^{0.7}\Lambda_0T_0^{0.7}}\frac{\alpha^{2.7}\dot E^{2.7}R_\star}{\dot M^{3.7}}
\end{equation}
For example, \citet{Lochhaas2017} derives $\beta_\mathrm{crit}$ for stellar winds in a massive star cluster from an instantaneous burst of star formation, valid in the few Myr before the first supernovae begin. In that case, the constant mass and energy deposition rates are dependent on the stellar mass of the star cluster, $\dot M=\beta\,10^{-3}M_\odot$/yr and $\dot E=\alpha\,10^{38.7}$ erg/s for $M_\star=10^5 M_\odot$ at solar metallicity. Note that in the case of an instantaneous burst of star formation driving stellar winds, $\beta$ is not the ratio of the wind mass to the star formation rate because there is no ongoing star formation after the initial burst. Instead, $\beta$ is the ratio of the injected wind mass to the expected mass loss rate from stellar winds and represents additional mass loading due to any material swept up within the cluster before the wind escapes to large scales. The stellar winds version of $\beta_\mathrm{crit}$ is
\begin{align}
&\beta_\mathrm{crit,SW}\simeq1.31 \nonumber \\
&\times\left(\frac{\alpha}{0.9}\right)^{0.73}\left(\frac{\mu}{0.6}\right)^{0.73}\left(\frac{R_\star}{1\,\mathrm{pc}}\right)^{0.27}\left(\frac{M_\star}{10^5 M_\odot}\right)^{-0.27}
\end{align}
and the stellar winds version of the maximum momentum of the escaping wind is
\begin{align}
    &\dot p_\mathrm{max,esc,SW}=2.69\ M_\odot\ \mathrm{yr}^{-1}\ \mathrm{km\ s}^{-1} \nonumber \\
    &\times\left(\frac{\alpha}{0.9}\right)^{0.86}\left(\frac{\mu}{0.6}\right)^{0.36}\left(\frac{R_\star}{1\ \mathrm{pc}}\right)^{0.14}\left(\frac{M_\star}{10^5 M_\odot}\right)^{0.86}.
\end{align}

For the remainder of this paper, we focus exclusively on the supernova-driven wind case, and only on the wind that escapes from the cluster, so we relabel $\dot p_\mathrm{max,esc}$ (equation~\ref{eq:pdot_max}) as simply $\dot p_\mathrm{max}$ and emphasize that $\dot p_\mathrm{max}$ always refers to the maximum momentum of only the wind that escapes. Table~\ref{tab:variables} lists the variables used in this paper and their meanings.

\begin{table}
    \centering
    \begin{tabular}{p{0.12\linewidth} | p{0.8\linewidth}}
        Variable name & Meaning\\
        \hline \\
        SFR & Star formation rate within wind-injection region (\S\ref{sec:betacrit}) \\
        $\alpha$ & Energy loading parameter within wind-injection region (\S\ref{sec:betacrit})\\
        $R_\mathrm{star}$ & Radius of wind-injection region (\S\ref{sec:betacrit})\\
        $\beta$ & Mass loading parameter \textit{within} wind-injection region (\S\ref{sec:betacrit})\\
        $\beta_\mathrm{esc}$ & Mass loading parameter of the \textit{escaping} wind (eq.~\ref{eq:beta_esc}, \S\ref{sec:betacrit})\\
        $\dot M_\mathrm{wind}$ & Mass injection rate \textit{within} the wind-injection region (eq.~\ref{eq:Mdot}, \S\ref{sec:betacrit})\\
        $\dot M_\mathrm{esc}$ & Mass loss rate of the \textit{escaping} wind (eq.~\ref{eq:Mdot_esc}, \S\ref{sec:betacrit})\\
        $\dot p_\mathrm{esc}$ & Momentum of the \textit{escaping} wind (eq.~\ref{eq:pdot}, \S\ref{sec:betacrit})\\
        $R_\mathrm{in,cool}$ & Radius inside which \textit{injected} mass cools and cannot escape the wind-injection region (Fig.~\ref{fig:Rcool}, \S\ref{sec:betacrit})\\
        $\beta_\mathrm{crit}$ & Mass loading \textit{within injection region} where injected mass just begins to cool and cannot escape the injection region (eqs.~\ref{eq:beta_crit},~\ref{eq:beta_crit_max}, \S\ref{sec:betacrit})\\
        $\dot p_\mathrm{max}$ & Maximum momentum of \textit{escaping} wind, corresponds to $\beta_\mathrm{crit}$ (eqs.~\ref{eq:pdotmax_betacrit},~\ref{eq:pdot_max}, \S\ref{sec:betacrit}) \\
        $R_\mathrm{out,cool}$ & Radius on large scales beyond which \textit{escaping} wind cools in a single phase (eq.~\ref{eq:Rcool_out}, \S\ref{sec:min})\\
        $\beta_\mathrm{esc,min}$ & Mass loading of the \textit{escaping} wind above which the wind can cool in a single-phase on large scales (eq.~\ref{eq:betaesc_min}, \S\ref{sec:min})\\
        $\beta_\mathrm{max}$ & Mass loading \textit{within injection region} below which the escaping wind can cool in a single-phase on large scales (Fig.~\ref{fig:betaesc}, \S\ref{sec:min})\\
        $\dot p_\mathrm{min}$ & Minimum momentum of \textit{escaping} wind under the requirement that it cools in a single phase on large scales, corresponds to $\beta_\mathrm{max}$ (\S\ref{sec:min})
    \end{tabular}
    \caption{The meanings of all variables used in this paper.}
    \label{tab:variables}
\end{table}

Equation~(\ref{eq:pdot_max}) is derived assuming the analytic approximation to the cooling function (equation~\ref{eq:Lambda}), but as we did for $\beta_\mathrm{crit}$, we can calculate $\dot p_\mathrm{max}$ using the full form of the cooling function tabulated by \citet{Wiersma2009}. We give the analytic forms in equations~(\ref{eq:beta_crit_max}) and~(\ref{eq:pdot_max}) to understand the scaling of $\beta_\mathrm{crit}$ and $\dot p_\mathrm{max}$ with the parameters of the problem, but for the remainder of this paper we use the numerically-calculated $\beta_\mathrm{crit}$ and $\dot p_\mathrm{max}$ with the full cooling function instead of the analytic forms. For most values of the parameters we explore, this leads to only minor differences in $\beta_\mathrm{crit}$ and $\dot p_\mathrm{max}$.

We test our $\beta_\mathrm{crit}$ and $\dot p_\mathrm{max}$ by implementing a wind-driving region in a 3D hydrodynamic simulation and measuring the wind's momentum outside of the wind-driving region for various values of $\beta$ using the Cholla code \citep{Schneider2015}, following the wind-driving model presented in \citet{Schneider2018a}. In a box with an initial low density ($n = 10^{-3}$ cm$^{-3}$) and high temperature ($T = 10^{6}$ K) background, we deposit mass and energy uniformly within a spherical region. The fiducial model is used to set all parameters of the driving region. Gas in the simulation is allowed to cool to a temperature floor of $T = 10^{4}$ K, assuming the piecewise-parabolic cooling function of \cite{Schneider2017}, which is a fit to a solar-metallicity cooling curve comparable to the \cite{Wiersma2009} cooling function. This mass and energy deposition drives an outward-moving spherical shock through the background material, after which the box settles into a steady-state wind that continually exits the box. Once the wind has reached a steady state, we measure the momentum of the escaping wind, $\dot p_\mathrm{esc}=\dot M_\mathrm{esc}v_\mathrm{wind}$, at a radius $r \gg R_\star$, for simulations with a range of beta values from $0.5 - 1.5$. As in the CC85 model, we do not include gravity.

\begin{figure*}
\begin{minipage}{175mm}
\centering
\includegraphics[width=\linewidth]{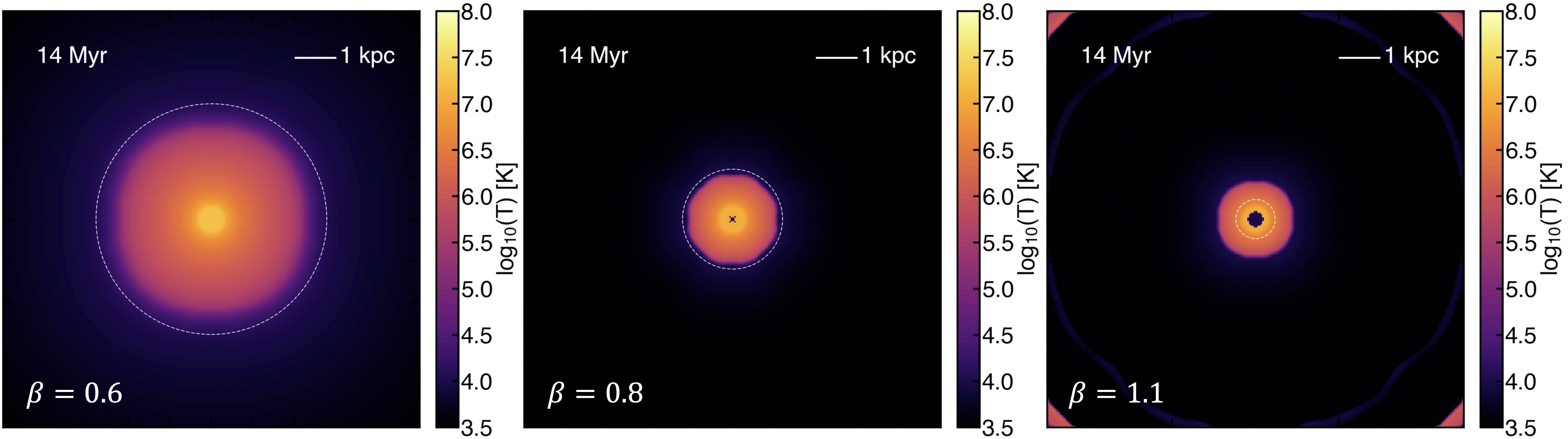}
\caption{Temperature slices through three examples from the suite of wind simulations, at a time when the wind has settled into a steady state. Each simulation is performed with the same set of fiducial parameters, but $\beta$ increases from left to right. The wind injection region is in the center of each slice, and a constant wind blows outward in all directions. As the wind expands outwards, it cools adiabatically and radiatively. The analytic cooling radius of the escaping wind given by \citep{Thompson2016b} is shown by the white dashed circle. As $\beta$ increases above $\beta_\mathrm{crit}$, the gas within the injection region begins to cool within $R_\mathrm{cool,in}$. This effect also causes the analytic cooling radius shown by the white dashed circle to underestimate the actual cooling radius of the escaping wind because it does not take into account the reduction in the mass outflow rate in the wind caused by cooling within the wind injection region.}
\label{fig:sims}
\end{minipage}
\end{figure*}

Figure~\ref{fig:sims} shows temperature slices for three examples from the suite of simulations using fiducial parameters. The wind injection region is located in the center. The interior of the wind injection region does not cool when $\beta<\beta_\mathrm{crit}$, which is shown in the left panel for $\beta=0.6$. The wind injection region is constant high temperature, and the wind decreases in temperature due to adiabatic expansion and single-phase radiative cooling as it flows away from the injection region. When $\beta\gtrsim\beta_\mathrm{crit}$, as in the middle panel where $\beta=0.8$, cooling has just set in in the very center of the injection region, and $R_\mathrm{cool,in}$ is small. When $\beta\gg\beta_\mathrm{crit}$, as in the right panel where $\beta=1.1$, most of the injection region cools and $R_\mathrm{cool,in}$ is large, greatly reducing the mass of the escaping wind.

\begin{figure}
\centering
\includegraphics[width=\linewidth]{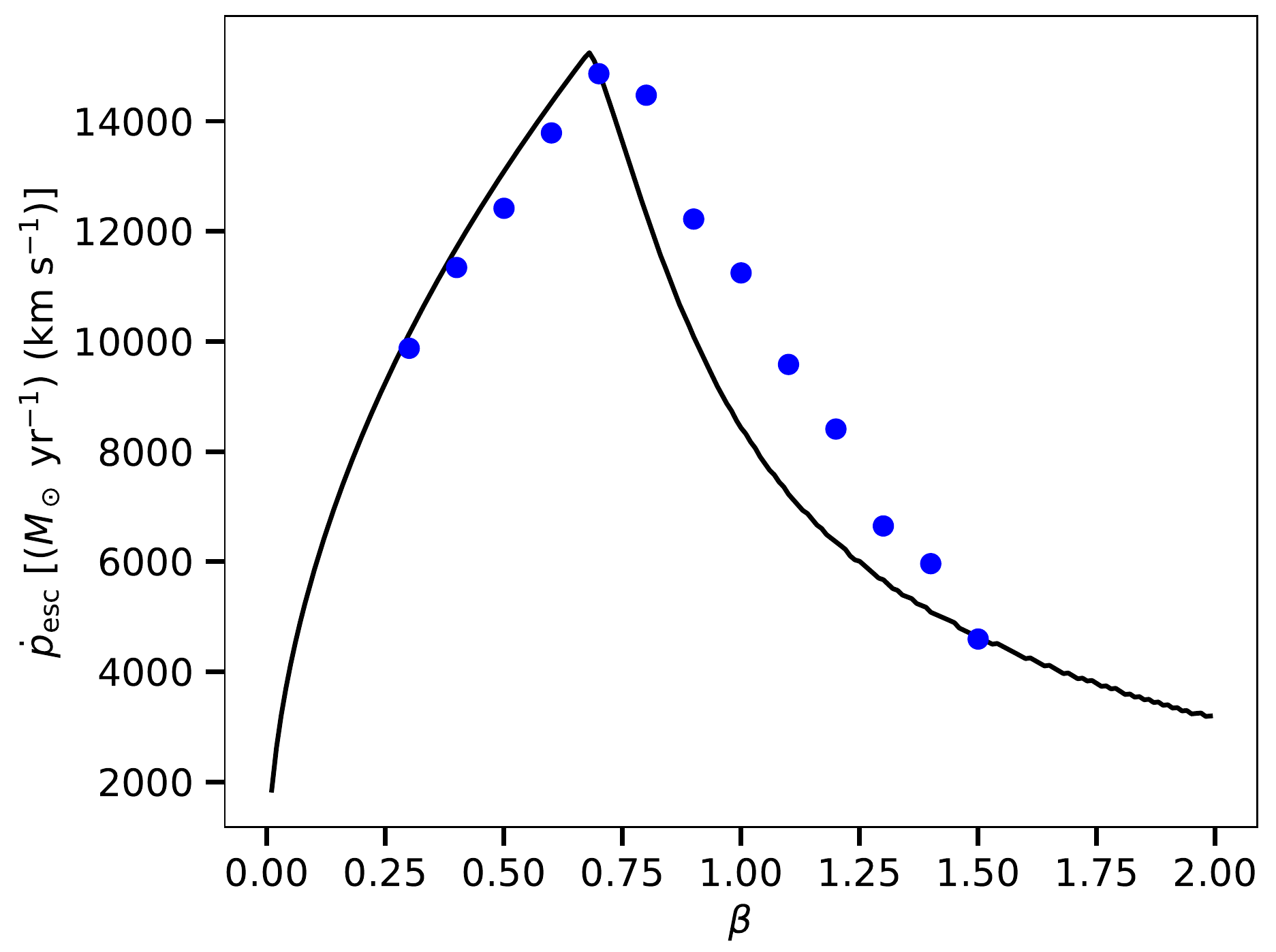}
\caption{The momentum of the escaping wind as a function of $\beta$ from the one-dimensional model is shown as the solid line and the momentum of the escaping wind from a suite of simulations implemented with different $\beta$s is shown as points. The maximum momentum occurs at $\beta=\beta_\mathrm{crit}$, and there is good agreement both in the value of $\dot p_\mathrm{max}$ and in $\beta_\mathrm{crit}$ between the one-dimensional model and the simulations. The slight disagreement at large $\beta$ is due to the assumption in the one-dimensional model that all wind deposited within $r<R_\mathrm{in,cool}$ is retained in the cluster, while in the simulation some of this wind escapes and adds to $\dot p_\mathrm{esc}$. Both the one-dimensional model and the simulations have $\alpha=0.9$, $\mu=0.6$, $R_\star=300$ pc, and $\mathrm{SFR}=20\ M_\odot$ yr$^{-1}$.}
\label{fig:pdot_simcompare}
\end{figure}

Figure~\ref{fig:pdot_simcompare} shows the momentum of the wind that escapes the wind-driving region as a function of $\beta$ from our one-dimensional model as the solid line and the wind momentum measured from our suite of simulations implemented with different $\beta$s as the points. There is good agreement between the two for small $\beta$ and $\dot p_\mathrm{max}$ has the same value $\approx1.5\times10^4\ M_\odot$ yr$^{-1}$ km s$^{-1}$ and occurs at the same $\beta_\mathrm{crit}\approx0.67$ in both, but there is a slight discrepancy in $\dot p_\mathrm{esc}$ for $\beta>\beta_\mathrm{crit}$. This is due to the assumption in the one-dimensional model that all wind deposited at $r<R_\mathrm{in,cool}$ is retained within the cluster, whereas in the simulations, the pressure gradient caused by the steep inner density profile and temperature floor results in some mass escaping the cooled region and contributing to $\dot{M}_\mathrm{esc}$, increasing $\dot p_\mathrm{esc}$. The disagreement is slight, so we consider this a validation of our one-dimensional model and continue with it for the remainder of the paper. Note that the difference in the value of $\dot p_\mathrm{max}\approx1.5\times10^4 M_\odot$ yr$^{-1}$ km s$^{-1}$ in Figure~\ref{fig:pdot_simcompare} from that given by equation~(\ref{eq:pdot_max}) $\dot p_\mathrm{max}=1.9\times10^4\ M_\odot$ yr$^{-1}$ km s$^{-1}$ is due to the analytic form of the cooling function giving a slightly different $\beta_\mathrm{crit}$, and thus slightly different $\dot p_\mathrm{max}$, as mentioned above.

\section{Minimum Momentum by Requiring Single-Phase Cooling on Large Scales}
\label{sec:min}

The temperature of a thermally-driven wind always decreases with distance as the wind expands adiabatically, regardless of its mass loading. A thermal supernova-driven wind launched by a starburst with a SFR of $10M_\odot$ yr$^{-1}$ and a low mass loading value of $\beta=0.2$ reaches $\sim10^4$ K by the time it has expanded 200 kpc from the galaxy \citep[see Figure~2 in][]{Thompson2016b}, as long as it does not interact with any other gas during its expansion. However, if the wind is more highly mass-loaded, radiative cooling can become efficient, and it is the combination of adiabatic expansion and cooling and radiative cooling that may cause the wind to reach $10^4$ K on smaller scales than hundreds of kpc. \citet{Thompson2016b} derived a minimum mass loading of the escaping wind ($\beta_\mathrm{esc}$ in our formalism) for the rapid, radiative cooling of the escaping wind outside the wind injection region, which we will call $\beta_\mathrm{esc,min}$. $\beta_\mathrm{esc,min}$ provides the minimum mass loading needed for the wind to cool on reasonable, sub-100 kpc scales within the halo.

We scale equation~(7) from \citet{Thompson2016b} to our parameters of interest and include the dependence on the molecular weight $\mu$ to find a minimum on $\beta_\mathrm{esc}$ that allows cooling of the escaping wind on large scales:
\begin{align}
&\beta_\mathrm{esc,min}\simeq0.34 \left(\frac{\alpha}{0.9}\right)^{0.636} \left(\frac{\mu}{0.6}\right)^{0.636} \left(\frac{R_\star}{300\ \mathrm{pc}}\right)^{0.364} \nonumber \\
&\times\left(\frac{\mathrm{SFR}}{20M_\odot\ \mathrm{yr}^{-1}}\right)^{-0.364} \label{eq:betaesc_min}
\end{align}
Figure~\ref{fig:betaesc} shows this minimum value for our fiducial parameters as the horizontal dashed curve. Wherever the solid curve, which indicates $\beta_\mathrm{esc}$, is larger than $\beta_\mathrm{esc,min}$, the wind is expected to radiatively cool on large scales in one phase.

The critical mass loading for cooling within the injection region derived in equation~(\ref{eq:beta_crit_max}), $\beta_\mathrm{crit}$, is larger than $\beta_\mathrm{esc,min}$, so if cooling has just set in within the injection region, it also occurs in the wind on large scales. Because $\beta_\mathrm{esc}$ decreases as $\beta$ increases above $\beta_\mathrm{crit}$, due to cooling within the wind driving region reducing the escaping wind mass, requiring $\beta_\mathrm{esc}>\beta_\mathrm{esc,min}$ translates to a maximum on $\beta$. We introduce a new parameter, $\beta_\mathrm{max}$, which describes the maximum $\beta$ that maintains a large enough $\dot M_\mathrm{esc}$ for the wind to single-phase cool radiatively on large scales. $\beta_\mathrm{max}$ is noted in Figure~\ref{fig:betaesc} with a point at the location where $\beta_\mathrm{esc}$ and $\beta_\mathrm{esc,min}$ cross. For our fiducial parameters, $\beta_\mathrm{max}\sim1.3$, so while there is still an escaping wind for $\beta\gtrsim1.3$, that wind remains hot because the escaping wind's mass loading ($\beta_{\rm esc}$) is sufficiently small that the flow does not become radiative on scales outside the injection region, assuming it does not mix with any mass outside the injection region. This implies there is a ``sweet spot" for large-scale single-phase cooling of thermally-driven galactic winds: the mass loading must be large enough for cooling on large scales but not so large that cooling within the injection region reduces the mass outflow rate so far that cooling no longer occurs on large scales. The minimum of the escaping wind mass, while still requiring large-scale cooling, is when $\beta=\beta_\mathrm{max}$, so $\dot p_\mathrm{esc}$ is minimized when $\beta=\beta_\mathrm{max}$. We use our numerical solution for $\beta_\mathrm{esc}$ as a function of $\beta$ (shown in Figure~\ref{fig:betaesc}) to find this minimum numerically, which we call $\dot p_\mathrm{min}$ (we have again dropped the subscript ``esc" as we did for $\dot p_\mathrm{max}$ because we are only concerned here with the momentum of the escaping wind). Table~\ref{tab:variables} lists the meanings of all variables used in this paper.

We also introduce an outer cooling radius, $R_\mathrm{out,cool}$, beyond which the hot wind is expected to become radiative and rapidly cool (if it does so at all). We reproduce equation~(6) from \citet{Thompson2016b} for the outer cooling radius for a given $\beta_\mathrm{esc}$, and include the dependence on mean particle weight $\mu$ from equation~(3) of \citet{Schneider2018}:
\begin{align}
&R_\mathrm{out,cool}\simeq 620\ \mathrm{pc} \nonumber \\
&\times\beta_\mathrm{esc}^{-2.92}\left(\frac{\alpha}{0.9}\right)^{2.13}\left(\frac{\mu}{0.6}\right)^{2.13}\left(\frac{R_\star}{300\ \mathrm{pc}}\right)^{1.79}\nonumber \\
&\times\left(\frac{\mathrm{SFR}}{20 M_\odot\ \mathrm{yr}^{-1}}\right)^{-0.789}. \label{eq:Rcool_out}
\end{align}
As cooling inside the wind-driving region reduces $\beta_\mathrm{esc}$, the outer cooling radius is located at larger radii. For example, if half of the volume of the wind-driving region cools, then the outer cooling radius becomes $R_\mathrm{out,cool}\simeq 4.7$ kpc for our fiducial parameters. If 90\% of the wind-driving region's volume cools, then $R_\mathrm{out,cool}\simeq 515$ kpc, effectively keeping the winds hot at any distance where they may be observed ``down the barrel." The effect of cooling within the wind injection region is to increase the single-phase cooling radius of the escaping wind to larger values than that derived in \citet{Thompson2016b}, so that the wind travels significantly further before cooling than it would if there were no cooling within the injection region. This effect can be seen in Figure~\ref{fig:sims}, where the white dashed circle shows the cooling radius calculated without taking this effect into account. In the right panel of the figure, the large $\beta$ causes the cooling radius of the outflow (where the temperature of the outflow drops to $T\sim10^4$ K) to be larger than expected.

We have derived a requirement for the hot winds to radiatively cool in a single phase, but recent works \citep{Gronke2018,Gronke2020,Kanjilal2020,Li2020,Schneider2020} present a picture in which hot winds can transfer their mass and/or momentum to cool material by mixing with some already-present cool clouds in the hot flow, such as lofted interstellar medium gas. Rather than the cool clouds being shredded and destroyed by the hot flow, the large clouds may survive, may be accelerated to the wind velocity, and may gain mass or momentum by mixing with and cooling from the hot flow at a constant rate. Recent works on turbulent mixing layers between hot and cool gas, while not specifically focusing on cold clouds embedded in a hot wind, come to a similar conclusion that the cold material grows by mass transfer from the hot phase \citep{Fielding2020,Tan2020}. Many of these works show the important parameter is the cooling time of the mixture of hot and cool gas, not the cooling time of the hot wind itself. In such a picture, the maximum $\dot p$ of hot winds derived in \S\ref{sec:betacrit} holds, as the hot wind cannot transfer more momentum to the cold cloud than it has. However, the minimum $\dot p$ for single-phase cooling derived in this section does not hold because large initial density perturbations in the wind, or interactions with cold clouds, can drive cooling even if the hot wind mass loading, $\beta_\mathrm{esc}$, is smaller than that formally required for the hot wind to cool, $\beta_\mathrm{esc,min}$. We note that our derived $\dot p_\mathrm{min}$ is not a strict lower limit for all models that produce cool gas from or within hot outflows; it is a lower limit only for models that assume single-phase cooling of hot gas \citep[e.g.,][]{Thompson2016b}

\section{Characteristic $\dot {\MakeLowercase{p}}$ of Single-Phase Cooling of Hot Winds}
\label{sec:char}

\begin{figure*}
\begin{minipage}{175mm}
\centering
\includegraphics[width=\linewidth]{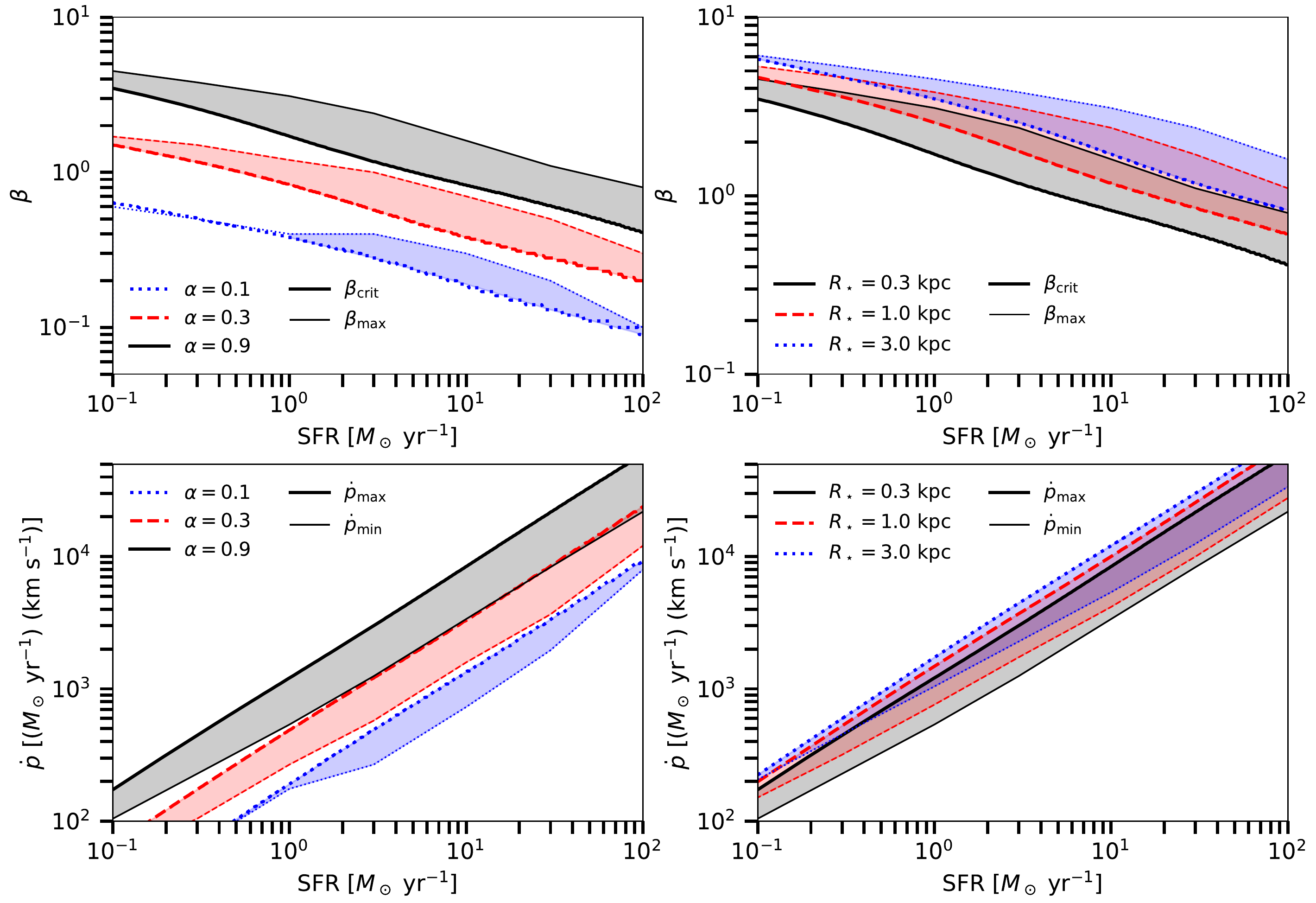}
\caption{The critical value of the mass loading, $\beta_\mathrm{crit}$, and the maximum value of the mass loading that allows cooling on large scales, $\beta_\mathrm{max}$, as functions of the SFR from the one-dimensional model in the top row, and the maximum and minimum values of the wind momentum, $\dot p_\mathrm{max}$ and $\dot p_\mathrm{min}$, as functions of the SFR in the bottom row. Note that $\beta_\mathrm{crit}$ corresponds to $\dot p_\mathrm{max}$ and $\beta_\mathrm{max}$ corresponds to $\dot p_\mathrm{min}$. Different curves in each panel show the effect of varying $\alpha$ (left column) and $R_\star$ (right column). In each panel, the $\beta_\mathrm{crit}$ or $\dot p_\mathrm{max}$ produced by our fiducial parameters is represented by the solid black curve. The fiducial parameters we assume are: $\alpha=0.9$, $\mu=0.6$, $R_\star=300$ pc.}
\label{fig:betacrit_pdotmax}
\end{minipage}
\end{figure*}

By combining the maximum (equation~\ref{eq:pdot_max}) and minimum (numerically solved from $\beta_\mathrm{esc}$, Figure~\ref{fig:betaesc}) on $\dot p$, we find a range of values of $\beta$ and $\dot p$ that hot winds can take in single-phase cooling models. These values are similar to each other, producing characteristic $\beta$ and $\dot p$ expected from single-phase cooling hot winds. We examine how these characteristic parameters vary with the parameters of the problem in Figure~\ref{fig:betacrit_pdotmax}. This figure shows $\beta_\mathrm{crit}$ (thick curves) and $\beta_\mathrm{max}$ (thin curves), with shading between them, in the top row and $\dot p_\mathrm{max}$ (thick curves) and $\dot p_\mathrm{min}$ (thin curves), again with shading between them, in the bottom row, both as a function of the star formation rate. Different curves in each panel show variation on the energy efficiency $\alpha$ (left column) and the size of the wind-driving region $R_\star$ (right column). Note that $\beta_\mathrm{crit}$ corresponds to $\dot p_\mathrm{max}$ (equation~\ref{eq:pdotmax_betacrit}) because there is a critical value of mass loading inside the wind injection region that produces a maximum escaping wind; $\beta_\mathrm{max}$ corresponds to $\dot p_\mathrm{min}$ (\S\ref{sec:min}) because mass loading above this maximum inside the wind injection region means the escaping wind can no longer radiatively cool in a single phase (Figure~\ref{fig:betaesc}). Thus, $\dot p_\mathrm{max}$ describes the maximum momentum of any thermally-driven wind, while $\dot p_\mathrm{min}$ describes the minimum wind momentum \textit{only} under the requirement that the wind radiatively cools in a single phase on large scales.

Larger values of $\dot M_\mathrm{wind}$ lead to denser winds that are more efficient at cooling. Because $\dot M_\mathrm{wind}\propto\mathrm{SFR}$, increasing the SFR leads to a denser wind-driving region that can cool more efficiently for a fixed $\beta$. Therefore, the value of $\beta$ that produces cooling within the wind-driving region decreases as SFR increases. For our fiducial model, $\beta_\mathrm{crit}\approx0.67$ for $\mathrm{SFR}=20\ M_\odot$ yr$^{-1}$.

The upper left panel shows that decreasing the efficiency with which stellar feedback energy couples to the outflow energy, parameterized by $\alpha$, decreases $\beta_\mathrm{crit}$. A lower $\alpha$ implies a cooler initial wind, as less feedback energy is thermalized in the gas. Because the peak of the radiative cooling curve is at lower temperatures, decreasing the temperature of the wind allows it to cool more efficiently, so the wind does not need a larger density from higher values of $\beta$ in order to promote cooling within the wind-driving region.

Increasing the radius of the wind injection region, $R_\star$, increases $\beta_\mathrm{crit}$ (upper right panel). A larger $R_\star$, for constant $\dot M_\mathrm{wind}$, produces a lower wind density within the wind-driving region. The lower wind density reduces the cooling efficiency and allows for larger values of $\beta$ before cooling sets in.

$\dot p_\mathrm{max}$ increases with increasing SFR (lower right panel) because the larger SFR contributes to more mass in the wind, and thus a larger momentum, even if the mass loading is smaller. However, the dependence of $\dot p_\mathrm{max}$ on the other parameters of $\alpha$ and $R_\star$ is the same as for $\beta_\mathrm{crit}$: more cooling within the injection region, as parameterized by a lower $\beta_\mathrm{crit}$, leads to a lower $\dot p_\mathrm{max}$. The effect of changing these parameters is smaller for $\dot p_\mathrm{max}$ than it is for $\beta_\mathrm{crit}$.

In general, the value of $\beta_\mathrm{max}$ is a factor of $\sim2$ higher than $\beta_\mathrm{crit}$ for a given set of parameters at a given SFR. However, in some cases, such as at low SFR when $\alpha=0.1$, $\beta_\mathrm{max}\approx\beta_\mathrm{crit}$. This indicates that while cooling within the wind driving region always leads to single-phase cooling within the wind on large scales, the large-scale cooling may be on scales so large as to say that the hot wind essentially does not cool on any scales of interest. In these cases, $\beta=\beta_\mathrm{crit}$ must be obtained in order for cooling to occur in the wind on large scales without interacting with cold clouds in the flow.

Although we assume a solar metallicity in our model, the radiative cooling curve is dependent on metallicity, as metal line cooling dominates the peak of the cooling curve. A complicating factor, and the reason why we leave metallicity out of this discussion, is that the mass loading of the wind affects its metallicity. Pure supernova ejecta has super-solar metallicity while the ISM, which provides the mass loading of the wind inside the injection region, in general has a lower metallicity. Therefore, the amount of ISM material mixed into the wind directly affects both $\beta$ and metallicity \citep{Chisholm2018} in a way that is not captured by either our analytic or our one-dimensional model.

\section{Comparison to Observations}
\label{sec:obs}

\begin{figure*}
\begin{minipage}{175mm}
\centering
\includegraphics[width=0.85\linewidth]{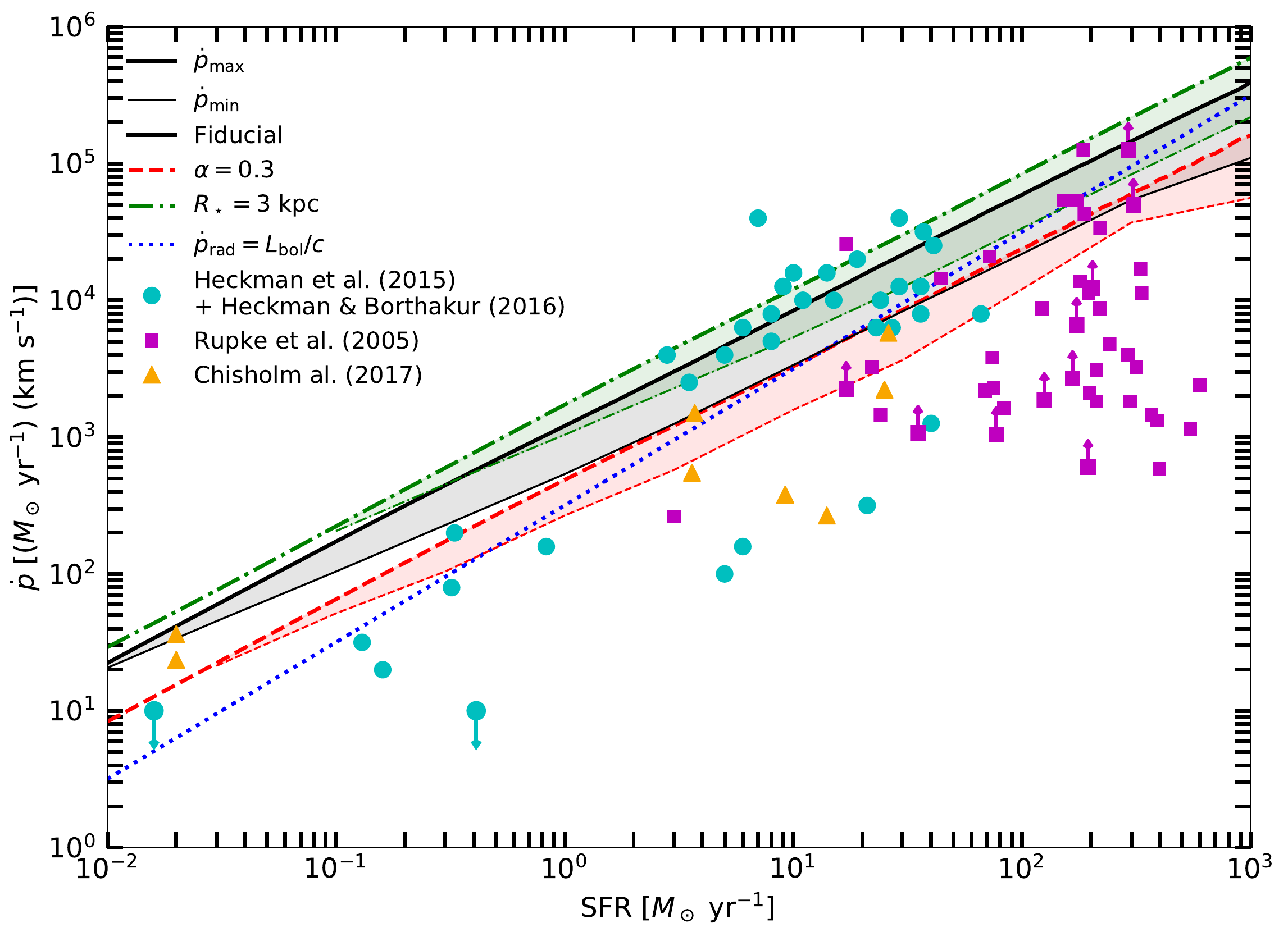}
\caption{The wind momentum rate $\dot p$ as a function of SFR. Solid black curves show our fiducial theoretical maxima (thick) and minima (thin) on the wind momentum, $\dot p_\mathrm{max}$ and $\dot p_\mathrm{min}$, with gray shading between. Other curves show $\dot p_\mathrm{max}$ (thick) and $\dot p_\mathrm{min}$ (thin), again with colored shading between, for variations on the fiducial model with parameter changes: $\alpha=0.3$ (red dashed) and $R_\star=3$ kpc (green dot-dashed). The expected $\dot p_\mathrm{rad}$ of radiation pressure driven winds in the single-scattering limit \citep{Murray2005} is shown as the blue dotted line. Our fiducial model has $\alpha=0.9$, $R_\star=300$ pc, and $\mu=0.6$. Cyan circles indicate the measured values of outflows from starforming galaxies from \citet{Heckman2015} and \citet{Heckman2016}, magenta squares indicate measured outflow momenta from \citet{Rupke2005}, and orange triangles indicate measured outflow momenta from \citet{Chisholm2017}. Arrows on points indicate upper or lower limits.}
\label{fig:obs_compare}
\end{minipage}
\end{figure*}

\citet{Rupke2005,Heckman2015,Heckman2016,Chisholm2017} observed the cool outflows from low-redshift galaxies in UV absorption against the galaxy starlight. Each study calculated the outflow velocity, outflow mass loss rate, and outflow momentum rate of a large number of galaxies. We compare the measured values of these outflows, $\dot p_\mathrm{out}$, with our theoretical maximum and minimum momenta, $\dot p_\mathrm{max}$ and $\dot p_\mathrm{min}$. Figure~\ref{fig:obs_compare} shows our fiducial model, as well as variations on the fiducial model by changing one parameter, compared to the observed wind values from these studies.

The majority ($75/88$) of the observed $\dot p_\mathrm{out}$ fall below the fiducial $\dot p_\mathrm{max}$ for various parameters and over a range in SFR, indicating the model is likely capturing the appropriate physics. However, only $17/88$ of the observational data lie between the fiducial $\dot p_\mathrm{min}$ and $\dot p_\mathrm{max}$. The maximum momentum is a stricter limit than the minimum in our framework, as the observed $\dot p_\mathrm{out}$ is dependent on the observed velocity of absorbing gas within the wind, which is subject to projection effects. Because velocities can only be measured along the line of sight, if gas is traveling at an angle to the line of sight, a velocity and momentum smaller than the true velocity of the flow may be observed. The observed momenta with $\dot p_\mathrm{out}<\dot p_\mathrm{min}$ are therefore not necessarily inconsistent with our model. In addition, as mentioned previously, our derived $\dot p_\mathrm{min}$ is that for single-phase cooling of hot winds only and neglects any interaction between hot winds and cold clouds outside of the wind-driving region (\S\ref{sec:min}).

Our theoretical limit $\dot p_\mathrm{max}$ is appropriate over a wide range in SFR; we see $\dot p_\mathrm{max}$ has a similar slope as the upper envelope of observed data. The measured $\dot p_\mathrm{out}$ is dependent on assumptions that are necessary in order to convert absorption lines into mass outflow rates, such as the opening angle of the flow and location of absorbing material along the line of sight toward the galaxy. For example, \citet{Heckman2015} assume a spherically symmetric flow (opening angle of $4\pi$) and that the absorbing material is at a distance from the starburst of about twice the radius of the starburst, which is also uncertain. A factor of two uncertainty in either the opening angle of the flow or the location of the absorbing material relative to the starburst produces a factor of two uncertainty in the measured mass outflow rate, and therefore a factor of two uncertainty in $\dot p_\mathrm{out}$. Within this uncertainty, nearly all the measured values of $\dot p_\mathrm{out}$ are consistent with being less than our theoretical maximum $\dot p_\mathrm{max}$.

Roughly $15\%$ ($13/88$) of the observed $\dot p_\mathrm{out}$ fall above the theoretical $\dot p_\mathrm{max}$ with $\alpha=0.9$ (solid black), if the assumptions about opening angle and position of absorbing material within the flow are correct. Prima facie, this implies that these observations cannot be tracing cool material that gained its momentum from the hot wind with our fiducial parameters. If the observational assumptions are correct, we can encompass more of the observed values by increasing $R_\star$ (green dot-dashed) to shift $\dot p_\mathrm{max}$ upward. This implies that if these observed cool outflows obtained their momentum purely from hot winds, then they must be escaping from a large wind-driving region in order to stay below the theoretical maximum momentum. \citet{Heckman2015} find the radii of the starburst regions in these galaxies are typically $\lesssim1$\,kpc, which is inconsistent with the model with $R_\star=3$ kpc shown as the green dot-dashed curve. Instead, to explain the high $\dot p_\mathrm{out}$ values, the observations could be tracing hot winds that shock on and sweep up additional material outside the wind injection region which obtains a ``momentum boost" before potentially cooling and being observed as cool outflows (see~\S\ref{sec:discussion}). In this case, the observed wind momentum is not constrained by $\dot p_\mathrm{max}$, if there is no error on the observational assumptions.

For those observed wind momenta that fall below the theoretical maximum, these outflows are consistent with a supernova-driven thermal wind. In these cases, if $\beta<\beta_\mathrm{max}$, the hot wind may be radiatively cooling in a single phase to produce the observations of cool outflows. If these cases have $\beta>\beta_\mathrm{max}$, where the wind mass outflow rate is too low for single-phase cooling on large scales, the observed outflows instead may be cool clouds accelerated by ram pressure from the hot flow, may be radiation pressure-driven winds, or mixing between hot and cool material may promote cooling and momentum transfer from hot wind to cool clouds within the outflow. The expected $\dot p_\mathrm{rad} = L_\mathrm{bol}/c$ for a radiation pressure driven wind from a starburst in the single-scattering limit \citep[equation~11 of][]{Murray2005}, where $L_\mathrm{bol}$ is the bolometric luminosity of the starburst for a given SFR, is plotted in Figure~\ref{fig:obs_compare} as the blue dotted line. While radiation pressure does not appear to produce large enough wind momenta to explain the largest $\dot p_\mathrm{out}$ data, this curve falls above those data with too small $\dot p_\mathrm{out}$ for single-phase radiatively cooling hot winds in our picture.

\begin{figure}
\centering
\includegraphics[width=\linewidth]{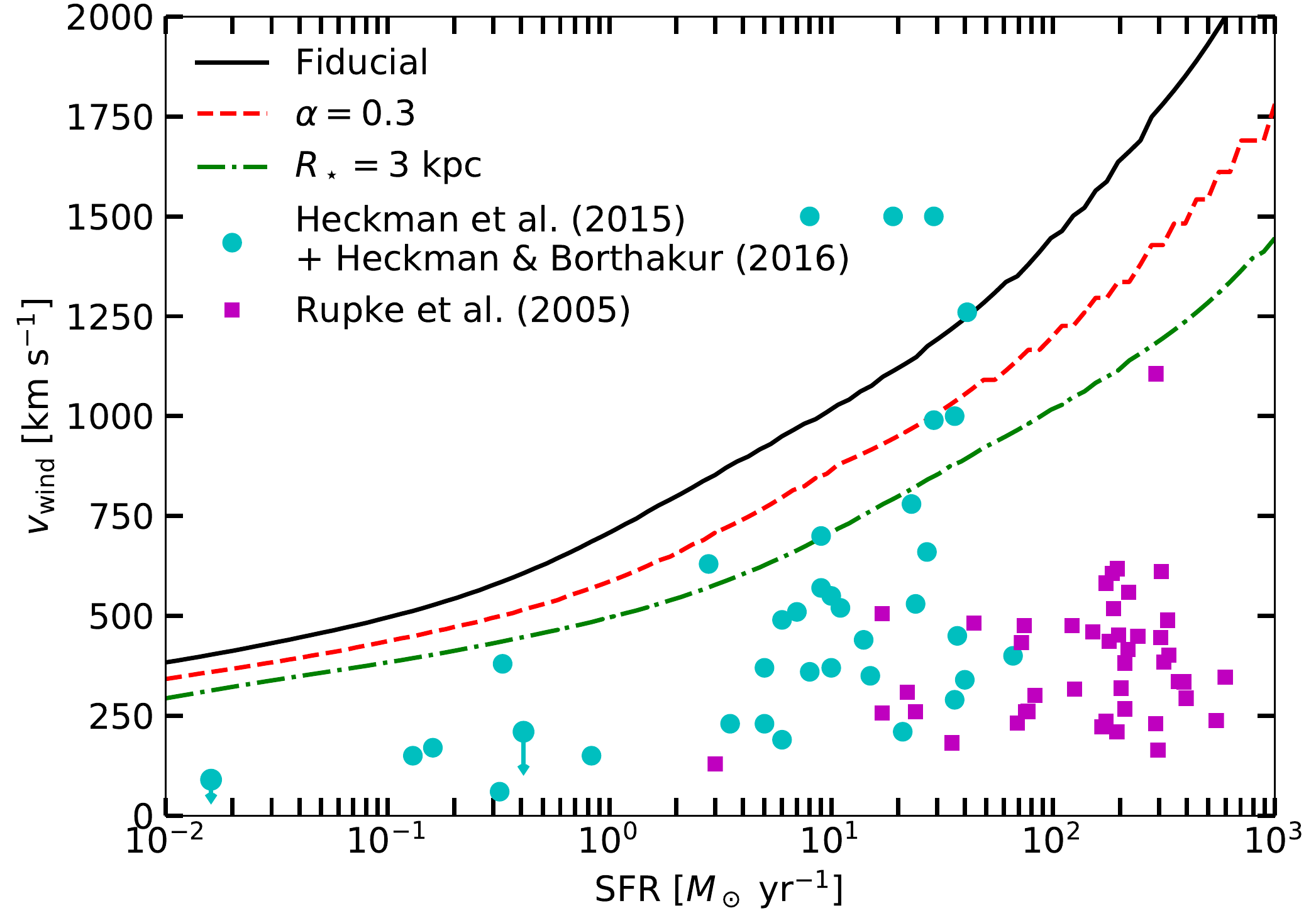}
\caption{The wind velocity at $\dot p_\mathrm{max}$ (and thus at $\beta_\mathrm{crit}$) as a function of SFR for fiducial parameters (black solid), with $\alpha=0.3$ (red dashed), or with $R_\star=3$ kpc (green dot-dashed). Measured wind velocities from \citet{Heckman2015} and \citet{Heckman2016} are plotted as light blue circles and wind velocities from \citet{Rupke2005} are plotted as magenta squares. \label{fig:vwind}}
\end{figure}

Figure~\ref{fig:vwind} shows the predicted wind outflow velocity when $\beta=\beta_\mathrm{crit}$ and $\dot p=\dot p_\mathrm{max}$ for the same sets of parameters as in Figure~\ref{fig:obs_compare}, and the measured $v_\mathrm{out}$ from the same studies \citep{Rupke2005,Heckman2015,Heckman2016}. The predicted wind velocity within the framework of the model may be higher or lower than what is plotted, if $\beta$ is higher or lower than $\beta_\mathrm{crit}$. \citet{Heckman2015} report the velocity centroid of absorption lines as the wind velocity, while \citet{Rupke2005} and \citet{Heckman2016} report the ``maximum" velocity, defined in \citet{Rupke2005} \citep{Heckman2016} as the velocity greater than the velocities of 90\% (98\%) of the detectable gas in the wind. The former method is column density weighted, which may not capture the fastest gas, while the latter method is subject to uncertainties due to the signal-to-noise of the spectrum and continuum placement. The predicted wind velocities are generally larger than the observed wind velocities, which supports the notion that there may be projection effects and spectral signal-to-noise effects reducing the observed velocities below their actual values and potentially explains the abundance of observed points with $\dot p_\mathrm{out}<\dot p_\mathrm{min}$. In addition, if the observed cool outflows are clouds in the process of being accelerated and mixed with the wind, they may not have yet reached the maximum velocity or momentum of the outflow at the time of observation, further explaining lower velocities than expected.

\section{Discussion}
\label{sec:discussion}

The observed somewhat too-high $\dot p_\mathrm{out}$ compared to the theoretical $\dot p_\mathrm{max}$ in 15\% of the observed galaxies indicates that these galaxies' outflow kinematics may not be purely described by the kinematics of the hot winds at launch, if the considerable measurement uncertainties are actually smaller than the factor of two enhancement between $\dot p_\mathrm{max}$ and the measurements. When a hot wind shocks on and sweeps up material outside of the wind-driving region, the swept-up material may experience a ``momentum boost" \citep{FaucherGiguere2012,Lochhaas2018} provided by the thermal energy produced by the wind shocking on this material. Winds that sweep up significant mass outside of the wind-driving region can deliver substantial boosts to the momentum of the swept material, from a factor of $\sim2$ for supernova-driven outflows up to a factor of $\sim20$ for AGN-driven outflows. The observed large values of $\dot p_\mathrm{out}$, if not driven by measurement uncertainties, may indicate that these outflows are interacting with substantial material outside of the wind-driving region that boosts the momentum of a wind-driven shell, which may then cool and be observed as cool outflows.

Many recent studies show how a hot wind can transfer mass and/or momentum to cool clouds \citep{Schneider2017,Gronke2018,Fielding2020,Gronke2020,Kanjilal2020,Li2020,Schneider2020,Sparre2020,Tan2020}. Cold clouds can survive in the hot flow so observations of outflows may be measuring the momentum transferred to the cool clouds rather than the momentum of the cooling hot phase itself. As the cool gas gains mass and/or momentum from the hot wind, its own momentum increases up to the maximum of the hot wind momentum. \citet{Vijayan2020} explicitly show that the cool phase (their ``warm") gains momentum from the hot phase in the outflow driven from a small patch of multiphase ISM. In this picture, the observed too-low cool gas momenta may be cool clouds that have not yet accelerated fully to the velocity of the hot flow, but are in the process of doing so. \citet{Schneider2020} found their cool outflows typically had factors of $\sim2-3$ less momentum than the hot wind, also consistent with this picture. The $\dot p_\mathrm{max}$ we derived for hot winds holds for these models and any other model in which hot material transfers momentum to the cold material, as the hot wind cannot transfer more momentum than it has.

However, our derived $\dot p_\mathrm{min}$ only holds for single-phase cooling of hot winds, as in the picture of \citet{Thompson2016b}. Contrary to the recent studies discussed above, several studies have shown that cool clouds embedded in a hot wind may not survive and will instead be ablated and mixed into the wind \citep{Klein1994,Cooper2009,Scannapieco2015,Bruggen2016,Ferrara2016,Schneider2017,Zhang2017}. In this picture, any cool clouds the flow impacts are heated to the high temperature of the wind, such that the hot wind itself must cool to produce observations of cool outflows. In this case, the observed wind momentum must adhere to both the maximum $\dot p_\mathrm{max}$ and $\dot p_\mathrm{min}$. The observed wind $\dot p$ that fall below $\dot p_\mathrm{min}$, if not beset by measurement uncertainties, projection effects, or low spectral signal-to-noise that affects the measurement of $v_\mathrm{wind}$, are therefore likely not described by a hot wind undergoing single-phase cooling as in \citet{Thompson2016b}.

Momentum transferal to cool clouds can explain small measured cool wind momenta, but a boost to the momentum requires that the cool clouds are swept up by the hot flow, shocked, momentum-boosted, and later cooled on larger scales. Other simulations of galactic winds that allow winds to evolve naturally from supernova-driven superbubbles \citep[i.e., do not tune the wind properties to the galaxy properties with an assumed scaling relation,][but see \citeauthor{Fielding2018} \citeyear{Fielding2018}]{Hopkins2012,Martizzi2015,Muratov2015,Martizzi2016,Fielding2017} find similar or slightly larger mass loading factors and wind velocities as measured in observational studies. The mass outflow rate in simulations is highly time-variable, typically peaking after a strong burst of star formation, consistent both with observations and the analytic model derived here that generally find a positive correlation between the wind momentum and SFR. \citet{Muratov2015} find the peak of the mass outflow rate can occur after the SFR burst has died down somewhat, so $\dot p_\mathrm{out}$ may be larger than would otherwise be expected for a given SFR, which may also help to explain the measurements with higher $\dot p_\mathrm{out}$ than the theoretical maximum.

The circumgalactic medium (CGM) is strongly impacted by galactic winds. Mass and metals in the CGM likely originate from a mix of SN ejecta and ISM material that is swept into galactic winds, so a limit on the mass loading and momentum of galactic winds also puts a limit on the mass and metal content of the CGM. The mass of the CGM cool phase alone is $10^{10}-10^{11}\ M_\odot$ \citep{Werk2014,Keeney2017,Prochaska2017} in $L^\star$ galaxies at $z\sim0.2$ and this phase has a metallicity of $\sim0.3\ Z_\odot$ \citep{Lehner2013,Prochaska2017}. If we assume that wind material has a metallicity of $\sim0.5\ Z_\odot$ to represent a mixture of SN ejecta and ISM material, and that any other material in the CGM is pristine gas, then the cool phase of the CGM must be made up of $60\%=6\times10^{9-10}\ M_\odot$ wind material. In order to produce $6\times10^{9-10}\ M_\odot$ of wind material in our fiducial model, a galaxy with an average SFR of $1\ M_\odot$ yr$^{-1}$ driving winds with maximal mass loading $\beta_\mathrm{crit}=2$ would require $3-30$ Gyr to populate the cool phase of the CGM with the observed amounts of mass and metals. Our fiducial model may explain the cool phase CGM observations for some galaxies, but additional mass loading of the wind outside the wind-driving region is likely necessary to match the larger mass estimates of the CGM.

\section{Summary}
\label{sec:summary}

We showed that when a mass-loaded wind is ejected from a wind-driving region, radiative cooling within the injection region can inhibit the escaping wind from the central area of the wind-driving region, producing a maximum wind momentum that occurs at a critical mass loading when cooling just sets in at the center of the wind-driving region. This maximum momentum of the escaping wind is not just a maximum on hot wind momentum rates, but is also a maximum on cool outflow momentum in any model where cool material gains momentum directly from the hot outflow, such as in single-phase radiatively cooling winds, ram pressure acceleration of cool clouds by a hot wind, or cool cloud growth or entrainment by mixing with a hot wind. In the single-phase radiatively cooling hot winds picture, there is also a minimum wind momentum required for the wind to cool on large scales, which obtains values similar to the maximum and thus produces a characteristic momentum for single-phase cooling hot winds. Our main findings are summarized as:
\begin{enumerate}
\item We derive the general form of a critical mass loading factor for a \citet{CC85} wind-driving region to radiatively cool in its interior, and apply parameters representing winds driven by either a constant supernova rate or stellar winds due to an instantaneous burst of star formation. For our fiducial parameters, the critical mass loading factor is $\beta_\mathrm{crit}\approx0.67$ (Figure~\ref{fig:betaesc}) for supernova-driven winds and $\beta_\mathrm{crit}\approx1.31$ for stellar winds in a massive star cluster. Using an analytic approximation to the cooling function, we find $\beta_\mathrm{crit}$ scales most strongly with the thermalization efficiency of wind energy and scales less strongly with the radius of the wind-driving region. It also scales inversely with the star formation rate (equation~\ref{eq:beta_crit_max}).
\item When $\beta<\beta_\mathrm{crit}$, the escaping wind mass and energy are equivalent to the injected wind mass and energy because no part of the injection region volume cools and retains wind material. When $\beta>\beta_\mathrm{crit}$, $\dot M_\mathrm{esc}/\dot M_\mathrm{wind}$ and $\dot E_\mathrm{esc}/\dot E_\mathrm{wind}$ decrease by the fraction of the injection region volume that radiatively cools (Figure~\ref{fig:Rcool}).
\item The critical mass loading implies a maximum momentum rate of hot winds (Figure~\ref{fig:pdot_simcompare}). For our fiducial parameters of a hot wind driven by SNe, we find $\dot p_\mathrm{max}\approx 1.9\times10^4\ M_\odot$ yr$^{-1}$ km s$^{-1}$. For stellar winds driven by an instantaneous burst of star formation in a massive cluster, we find $\dot p_\mathrm{max,SW}\approx 2.69\, M_\odot$ yr$^{-1}$ km s$^{-1}$. Using an analytic approximation to the cooling function, we find $\dot p_\mathrm{max}$ scales most strongly with the efficiency of wind energy and star formation rate, and scales less strongly with the size of the wind-driving region (equation~\ref{eq:pdot_max}).
\item Requiring single-phase cooling of the hot wind on large scales produces a maximum on the mass loading within the injection region and a minimum on the escaping wind momentum. In most cases, the maximum mass loading is $\sim2\beta_\mathrm{crit}$ (Figure~\ref{fig:betacrit_pdotmax}).
\item A comparison of the theoretical maximum and minimum momentum rates for supernova-driven winds to measured wind values by \citet{Rupke2005,Heckman2015,Heckman2016,Chisholm2017} shows that the majority of observed wind momenta fall below our theoretical maximum value, as predicted, but 15\% of observed winds have higher momenta than the theoretical maximum by a factor of $\sim2$ (Figure~\ref{fig:obs_compare}). Observational uncertainties could explain this discrepancy, but prima facie this implies that the observations may not be tracing cool outflows that obtained their momentum directly from the free-flowing hot wind, but rather there may be some mass loading outside the wind-driving region.
\item Over half of the observed wind momenta fall below the theoretical minimum value for single-phase cooling of the hot wind, and nearly all of the observed wind velocities fall below the predicted values (Figure~\ref{fig:vwind}), implying there are substantial projection effects reducing the observed wind velocities or that observations trace slower, cool clouds within the flow. The theoretical minimum momentum holds only when requiring single-phase cooling of the hot wind, as any model where cool material gains momentum from hot outflows can produce cool outflow momenta up to and including $\dot p_\mathrm{max}$.
\end{enumerate}

The theoretical limits on the mass and momentum rates of cooling hot winds derived in this paper are a first-principles explanation of a limit on the generation and impact of galactic winds that aligns well with observations. Cosmological simulations that generate galactic winds following a scaling relation between wind properties and galaxy properties should ensure that the resulting galactic winds are not more powerful than can be physically produced.

\section*{Acknowledgments}
CL thanks Max Gronke and Peng Oh for useful discussions. TAT is supported in part by NSF grant \#1516967 and NASA grant \#80NSSC18K0526. TAT acknowledges support from a Simons Foundation Fellowship and an IBM Einstein Fellowship from the Institute for Advanced Study, Princeton, while a portion of this work was completed. EES was supported in part by NASA through Hubble Fellowship grant \#HF-51397.001-A awarded by the Space Telescope Science Institute, which is operated by the Association of Universities for Research in Astronomy, Inc., for NASA, under contract NAS 5-26555.

\section*{Data Availability}
No new data were generated or analyzed in support of this research.

\end{document}